\documentclass[fleqn,usenatbib,useAMS]{mnras}

\usepackage{newtxtext,newtxmath}

\usepackage[T1]{fontenc}

\DeclareRobustCommand{\VAN}[3]{#2}
\let\VANthebibliography\thebibliography
\def\thebibliography{\DeclareRobustCommand{\VAN}[3]{##3}\VANthebibliography}

\usepackage{graphicx}	
\usepackage{amsmath}	
\usepackage{amssymb}	

\defcitealias{Thorngren2016}{T16}

\title[Heavy-metal Jupiters]{Heavy-metal Jupiters by major mergers:\\ 
metallicity vs. mass for giant planets
}

\author[S. Ginzburg and E. Chiang]{
Sivan Ginzburg$^{1}$\thanks{E-mail: ginzburg@berkeley.edu}\thanks{51 Pegasi b Fellow.}
and Eugene Chiang$^{1,2}$
\\
$^{1}$Department of Astronomy, University of California, Berkeley, CA 94720-3411, USA\\
$^{2}$Department of Earth and Planetary Science, University of California, Berkeley, CA 94720-4767, USA
}

\date{Accepted XXX. Received YYY; in original form ZZZ}

\pubyear{2020}

\begin{document}
\label{firstpage}
\pagerange{\pageref{firstpage}--\pageref{lastpage}}
\maketitle

\begin{abstract}
Some Jupiter-mass exoplanets contain $\sim$$100\, M_\oplus$ of metals, well above the $\sim$$10\, M_\oplus$ typically needed in a solid core to trigger giant planet formation by runaway gas accretion. We demonstrate that such `heavy-metal Jupiters' can result from planetary mergers near $\sim$10 au. Multiple cores accreting gas at runaway rates gravitationally perturb one another onto crossing orbits such that the average merger rate equals the gas accretion rate. Concurrent mergers and gas accretion implies the core mass scales with the total planet mass as $M_{\rm core} \propto M^{1/5}$ --- heavier planets harbour heavier cores, in agreement with the observed mass--metallicity relation. While the average gas giant merges about once to double its core, others may merge multiple times, as merger trees grow chaotically. We show that the dispersion of outcomes inherent in mergers can reproduce the large scatter in observed planet metallicities, assuming $3-30\, M_\oplus$ pre-runaway cores. Mergers potentially correlate metallicity, eccentricity, and spin.
\end{abstract}




\section{Introduction}

In the core accretion theory, a rocky/icy core accretes gas from the ambient circumstellar disc to form a giant planet \citep[e.g.][]{BodenheimerPollack86, Pollack1996}. Once the accreted atmosphere outweighs the core, the planet grows at an ever faster rate, as its Kelvin--Helmholtz cooling time now decreases with increasing mass \citep[e.g.][]{Ikoma2000}. The critical core mass needed to reach this runaway growth phase within the gas disc lifetime of a few Myr depends on the gas opacity and temperature, and is estimated to be $\sim$$10\,M_\oplus$ \citep[e.g.][]{PisoYoudin2014,Piso2015}.

\citet[][hereafter \citetalias{Thorngren2016}]{Thorngren2016} inferred the heavy element content of transiting giant planets by comparing their measured
masses and radii to evolutionary models.
The inference relies on the idea that at the same mass and age, metal-rich planets are denser and therefore smaller. They found that the planet's metallicity, normalized to that of its host star, decreases with
increasing planet mass. This trend is broadly consistent with core accretion theory --- planets with larger masses are dominated by their gas envelopes, which are presumably of stellar metallicity, and not by their heavy-element cores. However, the mass--metallicity relation in \citetalias{Thorngren2016} does not fit a model where all cores have the same mass 
$\sim$$10\,M_\oplus$: a constant core mass model
predicts a relation
that is somewhat too steep to explain the median
trend of the data, and more importantly,
it fails to explain the observed
large scatter in relative
metallicities, which can span an order of magnitude
or more at fixed mass.
Some Jovian planets have $\sim$$100\,M_\oplus$ of metals \citep[see also fig. 5 in][]{Barragan2018}.   

\citetalias{Thorngren2016} interpret 
this metal enrichment as due to late-stage
accretion of planetesimals, as has been
invoked to explain Jupiter's super-solar metallicity \citep[e.g.][]{Mousis2009}.
Late-stage accretion of solids is poorly 
constrained, depending on the unknown mass
and orbital distributions of 
small solid bodies, 
and their interaction
with residual disc gas. 
In \citetalias{Thorngren2016}'s model,
a planet accretes all of the solids, but 
not necessarily all of the gas,   
within a disc annulus centred on the planet
and spanning several Hill radii; the model posits
a spread of solid disc masses
to explain the spread of observed planet metallicities. 
The processes determining the solid accretion efficiency are not specified by
\citetalias{Thorngren2016}. 
Another form of late-stage accretion is explored by
\citet{Shibata2020}, who argue that orbital migration enables gas giants to capture tens of Earth masses in planetesimals.
However, the captured solid mass in their model does not seem to positively correlate with planet mass (their fig. 5), in contrast to the empirical relation found by \citetalias{Thorngren2016}. 

Here we consider how cores more massive than $10M_\oplus$ arise, not as a late-stage `afterthought', but as a natural outcome of an earlier phase of planet formation. Cores, including the terrestrial planets in our solar system, are thought to form within systems initially containing many protocores \citep[e.g.][]{KominamiIda2002}. These `oligarchies' \citep[e.g.][]{KokuboIda2012}
comprise bodies of comparable mass on nested orbits. Pairwise collisions\footnote{In the literature on galaxies, such collisions would be called `major mergers' --- a coalescence of objects of comparable mass.} between oligarchs successively double their mean mass and orbital spacing. In the solar system, mergers extend through the era of `giant impacts', one of which formed the Earth-Moon system.

In this paper we examine how mergers can play out at the same time that cores accrete gas from the nebula. We ask whether Jovian mass planets with especially massive cores (`heavy-metal Jupiters') can result from mergers.
Variations of this idea have been suggested in a similar context \citep[e.g.][]{Ikoma2006, Liu2015, Batygin2016}. A similar picture of giant impacts occurring contemporaneously with nebular gas accretion explains the metallicity diversity exhibited by super-Earths and sub-Neptunes, planets which never underwent runaway gas accretion \citep{Dawson2015,MacDonald2020}. 
We extend these ideas to gas-dominated planets to derive, for the first time, a quantitative mass--metallicity relation for merger products to compare against the \citetalias{Thorngren2016} data set. We are interested in reproducing both the observed mean trend and the scatter about the mean.

The remainder of this paper is organized as follows. In Section \ref{sec:mergers} we 
describe analytically 
how planets grow as they merge with each other and accrete gas from the circumstellar disc. In Section \ref{sec:core} we calculate 
how metallicity trends with mass, both in the mean and away from the mean, and compare with observations. 
A summary and discussion, including
potential observational tests, are given
in Section \ref{sec:discussion}. 

\section{Mergers of accreting planets}\label{sec:mergers}

\subsection{Before runaway}\label{sec:before}

We assume that planets grow from initially smaller rocky/icy protoplanets of equal mass $M$ that are uniformly separated from each other by a distance $\Delta a$ \citep[e.g.][]{KokuboIda2000}. Given enough time, the protoplanets perturb each other and merge. The time-scale to merge, computed from $N$-body simulations, is a steeply increasing function of $k\equiv \Delta a/R_{\rm H}$, where $R_{\rm H}\propto M^{1/3}$ is each protoplanet's Hill radius \citep[more precisely, the mutual Hill radius of each pair of neighbouring protoplanets; e.g.][]{Zhou2007}. In this section, we parametrize this time as 
\begin{equation}\label{eq:t_merge}
t_{\rm merge}\propto k^\alpha\propto\left(\frac{\Delta a}{M^{1/3}}\right)^\alpha
\end{equation}
with $\alpha\gg 1$; $\alpha=15$ fits the \citet{Zhou2007} results well in the relevant mass range for initially circular orbits. 
This simple power-law parametrization is useful for obtaining an analytical solution;
we will consider
more accurate expressions 
that account for non-zero initial eccentricity below.

Prior to runaway accretion, gas envelopes are limited to a small fraction of the total mass, and protoplanets grow almost exclusively by mergers. Merging adjacent protoplanet pairs doubles both $M$ and $\Delta a$, so $\Delta a\propto M$ as  the planets grow. From equation \eqref{eq:t_merge}, the merger time-scale increases as $t_{\rm merge}\propto M^{2\alpha/3}\simeq M^{10}$. 

While the protoplanets excite each other's eccentricities $e$, the gas disc damps eccentricities on a time-scale
\begin{equation}\label{eq:t_damp}
t_{\rm damp}= e/|\dot{e}| \sim \left(\frac{M_\star}{M}\right)\left(\frac{M_\star}{\Sigma_{\rm gas} a^2}\right) h^4 P_{\rm orb},  
\end{equation}
where $M_\star$ is the mass of the host star, $\Sigma_{\rm gas}$ is the disc gas surface density, $a$ is the orbital radius, $h$ is the disc aspect ratio, and $P_{\rm orb}\propto a^{3/2}$ is the orbital period. 
Equation \eqref{eq:t_damp} describes eccentricity damping from disc material located at the planet's first-order (as expanded in $e$) co-orbital Lindblad resonances, in the limit $e < h$ \citep[e.g.][]{Ward1988,Ward1989,Artymowicz1993,GoldriechSari2003,DuffellChiang2015}. The form of equation \eqref{eq:t_damp} can be derived heuristically
by considering dynamical friction exerted on the planet by co-orbiting gas \citep[see the appendix of][]{KominamiIda2002}.
The gas disc also causes planets to migrate, on a time-scale $t_{\rm mig} = a/|\dot{a}| \sim t_{\rm damp}h^{-2}\gg t_{\rm damp}$ \citep[e.g.][]{KleyNelson2012}. As we will see,
this is longer than any time-scale characterizing planetary growth (either by mergers or gas accretion),
and so migration can be safely ignored for our purpose of deriving 
the mass--metallicity relation. Migration can, however, transport fully-formed planets closer to their host stars.

\begin{figure}
	\includegraphics[width=\columnwidth]{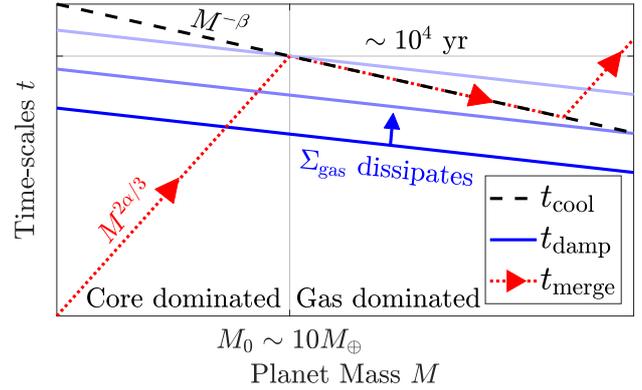}
	\caption{Schematic evolution of growing planets as they merge with each other and accrete gas from a depleting nebula. 
	We plot three time-scales: $t_{\rm merge}$ is the time for planets to perturb each other onto crossing orbits and merge, $t_{\rm damp}$ is the time it takes the gas disc to damp planetary eccentricities and thereby prevent merging, and $t_{\rm cool}$ is the planet's mass-doubling time through cooling-limited nebular accretion. All three times are equal
	at a `cross-over' mass $M_0$. At $M < M_0$,
	$t_{\rm merge}<t_{\rm cool}$ and planets grow
	by mergers and not by gas accretion; their masses are dominated by their solid cores (Section \ref{sec:before}). During this phase,
	the planets grow until
	$t_{\rm merge} = t_{\rm damp}$;
	the progression
	to higher masses is regulated by $t_{\rm damp}$,
	which shifts upward as the nebula dissipates
	(solid lines tinted blue).
	Once $M>M_0$, planets enter a runaway growth phase where they merge with each other and accrete nebular gas at the same rate $t_{\rm merge}=t_{\rm cool}$ (Section \ref{sec:during}). When runaway gas accretion ends (Section \ref{sec:after}), planets can continue to grow by mergers only (rightmost red arrow; this last phase is not modelled in our work).}
\label{fig:schematic}
\end{figure}

As long as $t_{\rm merge}<t_{\rm damp}$, protoplanets merge at the same rate as in a gas-free disc. Once $t_{\rm merge}\gtrsim t_{\rm damp}$, mergers halt, and the protoplanets stabilize on
non-crossing
orbits. As the gas disc gradually dissipates, $t_{\rm damp}\propto \Sigma_{\rm gas}^{-1}$ increases, enabling protoplanets to merge to progressively higher masses that satisfy $t_{\rm merge}=t_{\rm damp}$, as illustrated in Fig. \ref{fig:schematic} \citep[see also][]{KominamiIda2002}.

\subsection{During runaway}\label{sec:during}

In addition to mergers, planets that are embedded in a gas disc can also grow by accreting nebular gas. The accretion rate is initially
limited 
by the Kelvin--Helmholtz cooling time of the planet's gas envelope --- as the envelope cools and contracts, fresh gas from the nebula at large settles on top of the envelope and adds to its mass \citep{Ikoma2000, Lee2014, PisoYoudin2014}. Once the gas atmosphere outweighs the underlying solid core, cooling 
accelerates 
and the planet enters a runaway growth phase.

For nominal parameters
$a = 10\textrm{ au}$,
disc temperature
$T = 100\textrm{ K}$,
and disc aspect ratio $h= 0.06$ \citep{DAlessio98},
the cooling time is given by equation (19) of \citet{Ikoma2000}:
\begin{equation}\label{eq:t_cool}
t_{\rm cool}=\left(\frac{M}{M_\oplus}\right)^{-\beta}\left(\frac{\kappa}{10^{-2}\textrm{ cm}^2\textrm{ g}^{-1}}\right) {\rm Myr},   
\end{equation}
where, as before, $M$ is the total planet mass, and we have normalized the opacity $\kappa$ to that of
a dust-free gas. 
\citet{Ikoma2000} find a power-law slope of $\beta=2.5$ when $M\simeq M_{\rm core}$, where $M_{\rm core}$ is the underlying core mass. The same slope and normalization are found by \citet{LeeChiang2015} for their `dust-free and gas-rich beyond 1 au' case, after we plug $T = 100$ K into their equation (24) and equate the gas-to-core ratio to unity. \citet{PisoYoudin2014} find $\beta=2.4$. All these studies focused on the transition to runaway growth, when the planet's mass $M$ is still comparable to $M_{\rm core}$. Once the planet's mass is dominated by its gas atmosphere and $M\gg M_{\rm core}$, the slope flattens
to $\beta = 2$ \citep{GinzburgChiang2019a}. We adopt $\beta = 2$ for the remainder of Section \ref{sec:mergers}; 
in Section \ref{sec:core} we consider both $\beta=2$ and $\beta=5/2$.   

At some point the merging cores grow massive enough to initiate runaway gas accretion. This happens when the time to double a core's mass by merging with another core becomes longer than the time to double the mass by accreting a gas envelope (as given by equation \ref{eq:t_cool}). 
We denote the cross-over mass $M_0$ (see Fig. \ref{fig:schematic}): planet masses $M<M_0$ are dominated by their solid cores, whereas $M>M_0$ planets are gas dominated (for simplicity, we approximate $M=M_0$ planets as pure cores, although technically planets at cross-over have gas envelopes that constitute an order-unity fraction of their mass).
To evaluate $M_0$ ab initio, 
we would have to normalize the $t_{\rm merge}\propto M^{2\alpha/3}$ relation, i.e. specify the seed masses and orbital separations when giant impacts begin. Because these initial conditions are uncertain, we take a different approach and treat $M_0$ as a free parameter, with a nominal value of $M_0=10 M_\oplus$.
One can now `reverse' the calculation and estimate the separation $k = k_0$ when $M = M_0$. 
Equating $t_{\rm cool}$ from
equation \eqref{eq:t_cool} to $t_{\rm merge}$ as given by equations (3) and (4) of \citet{Zhou2007}, and evaluating the latter at 10 au and for an initial eccentricity $\tilde{e}\equiv 2ea/\Delta a\simeq 0.5$,
yields $k_0\simeq 5$, consistent with $N$-body simulations by \citet{KokuboIda2000}; see their fig.~7.
For $M_0=3M_\oplus$, $k_0 \simeq 7$, and for $M_0=30 M_\oplus$, $k_0 \simeq 4$.

Once $M>M_0$ and $t_{\rm cool}<t_{\rm merge}$, planets can significantly increase their mass without merging. According to equation \eqref{eq:t_merge}, merger-free 
growth increases $R_{\rm H}$ for the same separation $\Delta a$, shortening $t_{\rm merge}\propto M^{-\alpha/3}$. If $\beta<\alpha/3$ (true for our nominal parameters), then $t_{\rm merge}$ decreases below $t_{\rm cool}$, contradicting our assumption of growth without mergers. We conclude that the system is driven towards concurrent runaway gas accretion and mergers, satisfying $t_{\rm merge}(M,\Delta a)=t_{\rm cool}(M)$ as depicted in Fig. \ref{fig:schematic} 
(dashed black line overlaid by the middle red arrow). 
We solve this equation to find how, on average, the separation between adjacent protoplanets changes as they simultaneously
accrete gas and merge:
\begin{equation}\label{eq:delta_a}
\Delta a(M)\propto M^{1/3-\beta/\alpha}\simeq M^{1/5}   
\end{equation}
where we have substituted our nominal $\alpha=15$ and $\beta=2$. This result lies between growth without mergers ($\Delta a\propto M^0$) and growth by mergers only ($\Delta a\propto M^1$; Section \ref{sec:before}). 

As gas-dominated planets merge, so do their cores. The increasing separation $\Delta a(M)$ 
scales directly with the number of mergers and thus
with the number of merged cores, each of mass $M_0$:
\begin{equation}\label{eq:m_core}
\frac{M_{\rm core}(M)}{M_0}=\frac{\Delta a(M)}{\Delta a(M_0)}=\left(\frac{M}{M_0}\right)^{1/3-\beta/\alpha}\simeq\left(\frac{M}{M_0}\right)^{1/5} 
\end{equation}
where 
$M_{\rm core}(M_0)\equiv M_0$. 
Although the core mass grows with successive mergers, 
the core's fraction of the total mass decreases as
\begin{equation}\label{eq:m_core_frac}
\frac{M_{\rm core}}{M}=\left(\frac{M}{M_0}\right)^{-2/3-\beta/\alpha}\simeq \left(\frac{M}{M_0}\right)^{-4/5}.    
\end{equation}
It is instructive to compare equations \eqref{eq:m_core} and \eqref{eq:m_core_frac} to 
conventional 
core-accretion theory, in which planets accrete their gas in isolation without merging. In the isolated scenario, $M_{\rm core}$ is constant and $M_{\rm core}/M\propto M^{-1}$ --- steeper than the $M^{-1/2}$ proportionality found empirically by \citetalias{Thorngren2016}. 
With mergers, $M_{\rm core}/M\propto M^{-4/5}$ --- closer to the observed relation, but still not shallow enough at face value. 
In fact, if the original critical cores have a mass $M_0\sim 10 M_\oplus$, equation \eqref{eq:m_core} indicates that Jovian planets with $M \sim 300 M_\oplus$ experienced only one merger on average, 
doubling the core mass $(M/M_0)^{1/5}\simeq 2$. These statements
apply only in the mean, however; both the observations, and the merger process that we propose underlies them, have large scatter. The more telling test of the theory will
be to account for this scatter, a topic 
we address in Section \ref{sec:scatter}.

Another revealing exercise is to take 
equations \eqref{eq:m_core} and \eqref{eq:m_core_frac} 
in the limit $\alpha\gg\beta$, i.e. to consider a very steep dependence of $t_{\rm merge}$ on $k$, the number of Hill radii separating adjacent protoplanets. In this limit, concurrent gas accretion and mergers tend to keep the system at a constant $k$: if $k$ becomes too high, mergers stop and planets grow by runaway gas accretion, decreasing $k$; if $k$ becomes too low, planets merge before they accrete much gas from the nebula, increasing $k$ (as in Section \ref{sec:before}). Equations
\eqref{eq:m_core} and \eqref{eq:m_core_frac} then reduce to
$M_{\rm core}\propto M^{1/3}$ and $M_{\rm core}/M\propto M^{-2/3}$. These resemble relations posited by \citetalias{Thorngren2016}, who take $M_z\propto M^{1/3}$ for the mass in metals accreted from the disc, during or after runaway gas accretion, and find by extension that the planet metallicity $M_z/M\propto M^{-2/3}$. 
The resemblance, which stems
from the use of $R_{\rm H}$ in our theory and theirs, 
is only coincidental, as $R_{\rm H}$ plays
a different role between the two pictures:
\citetalias{Thorngren2016} argue that $M_z \propto M^{1/3}$ based on the idea that the
Hill radius $R_{\rm H} \propto M^{1/3}$ 
sets the size of the disc annulus
from which the planet accretes solids,
whereas we find $M_{\rm core} \propto M^{1/3}$
because $R_{\rm H}$ is used to determine the
time-scale over which the system is unstable to mergers
(equation \ref{eq:t_merge}). 
In any case, the $\alpha\to\infty$ limit captures the main 
ideas of our model and provides 
intuition. It can also accommodate other prescriptions for $t_{\rm merge}$. 
For example, resonance overlap theory predicts that orbit stability is sensitive to $\Delta a/M^{2/7}$ \citep{Wisdom80,Duncan89,Deck2013}, rather than $\Delta a/M^{1/3}$. Then the $\alpha\to\infty$ limit yields $M_{\rm core}\propto M^{2/7}$. 
\citet{YalinewichPetrovich2020} derive
another formulation for $t_{\rm merge}$
that can also be accommodated.

So far we have assumed that the planets in a multi-planet system remain comparable in mass as they grow. While this assumption may hold prior to runaway gas accretion, it breaks down once the first planet reaches the cross-over mass $M_0$. 
In the most extreme scenario, only one core 
runs away, growing to $M\gg M_0$, while
its companions remain slightly sub-critical at $\sim$$M_0$. In this case, the runaway planet's core mass $M_{\rm core}(M)$ is smaller, when compared to the equal mass case, because the planet has to reach a higher mass $M$ to merge with its low-mass neighbours.
Quantitatively, the mutual Hill radius $R_{\rm H}$
of the planet and its neighbour is smaller by a factor of 
$[(M_0+M)/(2M)]^{1/3}$, or about $(1/2)^{1/3} \simeq 0.8$ for
$M \gg M_0$, as compared to the equal mass scenario. From the condition $t_{\rm merge}(\Delta a/R_{\rm H})=t_{\rm cool}(M)$, $M_{\rm core}(M)\propto\Delta a(M)\propto R_{\rm H}(M) M^{-\beta/\alpha} \propto M^{1/5}$ as before, but with a numerical coefficient that is smaller by a factor of 0.8 
(arising from $R_{\rm H}$).
Because this correction is much smaller than the scatter in the merger process (Section \ref{sec:scatter}), we drop it for the remainder of our study.
We note, however, that the function $t_{\rm merge}(k)$ itself may change for unequal-mass planets \citep{PuWu2015}. 

The assumption that planet--planet perturbations lead to mergers rather than ejections is justified by showing that the escape velocity from the planets is smaller than that from the star. During runaway gas accretion, planets are puffy --- their boundaries extend to the Bondi radius, which implies their surface escape velocity is of order the disc's sound speed \citep[e.g. fig.1 in][]{GinzburgChiang2019b}. This is smaller than the escape velocity from the star, which is comparable to the Keplerian orbital velocity, 
by a factor of order the disc aspect ratio $h\ll1$. Planet scatterings are therefore unlikely to eject planets from the system.

\subsection{Mass budget during runaway}\label{sec:budget} 

As Fig. \ref{fig:schematic} illustrates, the gas surface density $\Sigma_{\rm gas}$ must be low enough for giant impacts to produce planets of mass $M_0$. Once this critical mass is reached, the merger and cooling time-scales are both shorter than the eccentricity damping time, such that growth can proceed without further reduction of $\Sigma_{\rm gas}$. 
We now check whether this surface density, which sets the gas mass available for accretion onto planets, is sufficient to form gas giants with masses of 0.1--10 $M_{\rm Jup}$, where $M_{\rm Jup}$ is Jupiter's mass. 
By equating equations \eqref{eq:t_damp} and \eqref{eq:t_cool} and substituting our nominal $\kappa=10^{-2} \textrm{ cm}^2\textrm{ g}^{-1}$, $a=10\textrm{ au}$, $h=0.06$, and $M_\star = M_\odot$ (a solar mass star),
we find that the local gas mass $\Sigma_{\rm gas}a^2$ 
is given by
\begin{equation}\label{eq:gas_mass}
\begin{split}
\frac{\Sigma_{\rm gas}a^2}{M_\star}&=4\times 10^{-10}\frac{M_\star}{M_0}\left(\frac{M_0}{M_\oplus}\right)^\beta\left(\frac{a}{10 \textrm{ au}}\right)^{37/14}\\
&\simeq \left(\frac{a}{10 \textrm{ au}}\right)^{37/14}\times\begin{cases}
1\times 10^{-3} & {\rm for}\,\, \beta=2\\
4\times 10^{-3} & {\rm for}\,\, \beta=5/2
\end{cases}
\end{split}
\end{equation}
when the critical mass $M_0 = 10 M_\oplus$ is reached.
Equivalently, 
\begin{subequations}\label{sigma_gas}
\begin{align}
&\Sigma_{\rm gas}\simeq 100\textrm{ g cm}^{-2}\left(\frac{M_0}{10M_\oplus}\right)\left(\frac{a}{10\textrm{ au}} 
\right)^{9/14}  \enspace &{\rm for}\,\, \beta=2\\
&\Sigma_{\rm gas}\simeq 400\textrm{ g cm}^{-2}\left(\frac{M_0}{10M_\oplus}\right)^{3/2}\left(\frac{a}{10\textrm{ au}} 
\right)^{9/14}  \enspace &{\rm for}\,\, \beta=5/2
\end{align}
\end{subequations}
which is a few times more massive than the minimum-mass solar nebula (MMSN) at $a=10\textrm{ au}$  \citep[e.g.][from which we adopt $h\propto a^{2/7}$]{ChiangYoudin2010}.

Equation \eqref{eq:gas_mass} indicates that 
the disc has enough gas left to form gas giants at $a \gtrsim 10$ au. It also implies that massive discs have to deplete to Jovian masses before they can form gas giants --- Jupiter's mass emerges as a natural scale for the outcome of runaway accretion,
irrespective of the original disc mass
\citep[cf.][]{Tanaka2020,Rosenthal2020}.
At the same time,
equation \eqref{eq:gas_mass} also implies
that gas densities may be too low at $a < 10$ au
to form gas giants there. 
Eccentricity damping is significantly
faster closer to the star: according to equation \eqref{eq:t_damp}, $t_{\rm damp}\propto (\Sigma_{\rm gas}a^2)^{-1}a^{37/14}$ for $h\propto a^{2/7}$, 
implying that gas densities
have to deplete to prohibitively
low values for cores to emerge by mergers.
Furthermore, the cooling time is 
longer closer to the star, requiring $\Sigma_{\rm gas}$ to decrease even more to satisfy $t_{\rm damp}=t_{\rm cool}$ at $a<10\textrm{ au}$. 
Taking $t_{\rm cool}\propto a^{-15/14}$ \citep{PisoYoudin2014} changes our results to $\Sigma_{\rm gas}a^2\propto a^{26/7}$ and $\Sigma_{\rm gas}\propto a^{12/7}$.
The fact that disc gas densities must be especially
low close to the star before $t_{\rm damp} = t_{\rm cool}$
was used by \citet{LeeChiang2016}
to explain the prevalence of super-Earths/sub-Neptunes, which have low atmospheric mass fractions, at $a < 1$ au.
Here it suggests that gas giants can only 
form at $a > 10$ au and must subsequently migrate
inward to $a < 1$ au where they are currently observed.
This is consistent with 
the ordering of time-scales $t_{\rm mig} > t_{\rm damp} > t_{\rm merge} \sim t_{\rm cool}$ characterizing
the phase of concurrent mergers and gas accretion.

We also gauge whether the disc has enough solid mass to form cores of mass $M_0$. If each core accretes all the solids that are closer to it than to  adjacent cores,
then $M_0=2\upi a k_0 R_{\rm H}\Sigma_{\rm solid}$, where $\Sigma_{\rm solid}$ is the solid surface density. Taking the mutual Hill radius $R_{\rm H}=[(2/3)(M_0/M_\star)]^{1/3}a$, and assuming $k_0=5$ (Section \ref{sec:during}), we find
\begin{equation}\label{eq:sigma_solid}
\begin{split}
\Sigma_{\rm solid}=&\left(\frac{3}{2}\right)^{1/3}\frac{M_0^{2/3}M_\star^{1/3}}{2\upi k_0 a^2}
\\
\simeq &\,3\textrm{ g cm}^{-2}\left(\frac{M_0}{10M_\oplus}\right)^{2/3}\left(\frac{a}{10\textrm{ au}}\right)^{-2},  
\end{split}
\end{equation}
which is about 3 times more massive than the MMSN at $a=10\textrm{ au}$. 

\subsection{After runaway}\label{sec:after}

As shown in Fig. \ref{fig:schematic}, mergers and cooling-limited accretion unfold on ever shorter times-scales as planets grow in mass beyond $M_0$. At some point, however, runaway gas accretion ends. 
It may stop simply because there is no gas left in the disc (Section \ref{sec:budget}).
Or it may stop because the planets open gaps in the
gas disc; for massive enough planets, the bottleneck
for gas accretion
is no longer the cooling rate of the atmosphere, but rather the hydrodynamic rate at which gas flows from the disc through the gap to the planet.
\citet{GinzburgChiang2019a} find that in low-viscosity discs, the planet's growth time eventually changes from $M/\dot{M}=t_{\rm cool}\propto M^{-2}$ to $M/\dot{M}\propto M^{15}$, halting runaway growth (their fig. 1; see also \citealt{Rosenthal2020}). 

In any case, as gas accretion slows,
mergers make a comeback as the dominant growth mode.
But the time-scale for mergers unassisted by
gas accretion is prohibitively long,
lengthening as 
$t_{\rm merge}\propto M^{10}$
(Section \ref{sec:before} and the rightmost red arrow of Fig. \ref{fig:schematic}).
Accordingly, we assume that planets
reach their final mass 
around the time
that runaway growth ends (i.e.
near
the juncture of the middle and rightmost red arrows
in Fig. \ref{fig:schematic}).

\section{Core mass and metallicity}\label{sec:core}

In Section \ref{sec:during} we calculated how the heavy-element core grows through mergers of neighbouring planets 
using a simple power-law parametrization of the merger rate. In this section we refine our analysis by incorporating the results of published
$N$-body simulations, and compare our results to the observed mass-metallicity relation.   

\begin{figure}
	\includegraphics[width=\columnwidth]{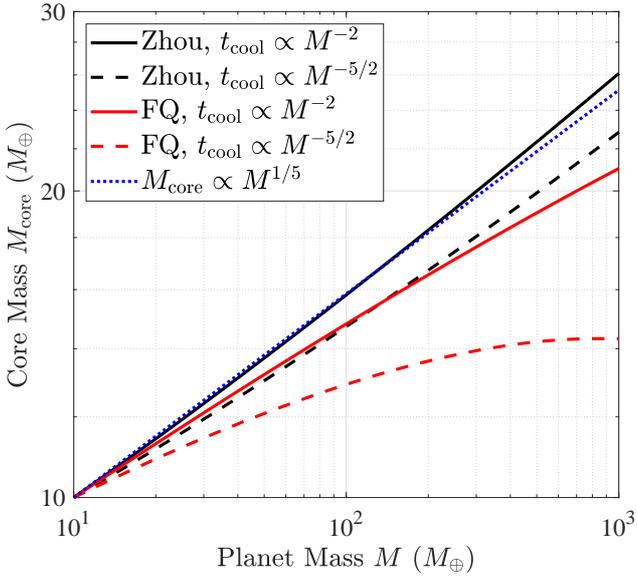}
	\caption{Growth of planetary cores as a result of
	concurrent mergers and nebular gas accretion.
	The initial planets are pure cores (without gas) weighing $M_0 = 10 M_\oplus$, the critical cross-over mass when runaway gas accretion becomes competitive with mergers (see Section \ref{sec:during}).
	The core mass $M_{\rm core}$ as a function
	of the total mass $M$ (core + envelope)
	is computed by equating the time-scale for mergers,
	$t_{\rm merge}$, with the envelope cooling/accretion
	time-scale $t_{\rm cool}$. As annotated, we draw $t_{\rm merge}$
	from either \citet[][`Zhou']{Zhou2007} or \citet[][`FQ']{FaberQuillen2007}, and $t_{\rm cool}$ from
	equation (\ref{eq:t_cool}) using either $\beta=5/2$
    (more accurate for $M \simeq M_{\rm core}$) 
    or $\beta =2$ (better for $M \gg M_{\rm core}$).  
    By the time planets become gas giants, they are likely to have undergone 
    at least one merger
    which doubles 
    the core mass (at high $M$,
    the dashed red
    line is not realistic since
    it uses $\beta = 5/2$ when it
    should use $\beta = 2$).
    The $M_{\rm core}\propto M^{1/5}$ approximation (dotted blue line), derived in Section \ref{sec:during}, follows closely our nominal model
    (solid black line) used to compute Fig. \ref{fig:metallicity}.}
\label{fig:core}
\end{figure}

In Fig. \ref{fig:core} we present $M_{\rm core}(M)$ for our nominal critical mass $M_0=10 M_\oplus$ and for two choices of the cooling power-law index $\beta$. 
As in equations \eqref{eq:delta_a} and \eqref{eq:m_core}, $\Delta a(M)$ is given by the condition $t_{\rm merge}(M,\Delta a)=t_{\rm cool}(M)$
and the core mass is given by $M_{\rm core}(M)/M_0=\Delta a(M)/\Delta a(M_0)$.
The only difference between the present calculation and the one in Section \ref{sec:during} is that here $t_{\rm merge}$ is not given by the power-law equation \eqref{eq:t_merge}, but rather by empirical fits to $N$-body simulations, either by \citet{Zhou2007} or \citet{FaberQuillen2007}.\footnote{Alternative $N$-body simulations
considered by \citet[][their fig. 3]{PuWu2015}  
yield moderately shorter merger time-scales compared to \citet{Zhou2007}, by up to a factor of 10 (at fixed $k$ 
for instability times $< 10^{6.5} \,\times$
the period of the innermost planet).} Whereas the \citet{Zhou2007} fit is essentially a (mass-dependent) power law in $\Delta a/M^{1/3}$, \citet{FaberQuillen2007} fit an exponential
in $\Delta a/M^{1/4}$ (see also \citealt{YalinewichPetrovich2020}). We scale both relations to the orbital period at 10 au, and assume zero initial eccentricity when using the fit of \citet[][]{Zhou2007}. 
Since merger times are shorter for eccentric orbits, the core masses we compute in Fig. \ref{fig:core} are lower bounds; we discuss eccentric orbits in Section \ref{sec:scatter}. 

Our nominal model for the remainder of this paper uses $t_{\rm merge}$ from \citet{Zhou2007} and $t_{\rm cool}\propto M^{-2}$;  shown by a solid black line in Fig. \ref{fig:core},
it is reproduced well by our analytical equation \eqref{eq:m_core} (dotted blue line).
Other model choices give similar results: they imply that planets undergo between one and two mergers, which 
roughly
double 
their core mass by the time they become gas giants (the saturation of the dashed red line in Fig. \ref{fig:core} is
not realistic; for
$M\gg M_{\rm core}$,
$\beta = 5/2$ is not
as accurate as
$\beta=2$).

We use the calculated core masses $M_{\rm core}(M)$ to evaluate the planet's bulk metallicity, defined as $Z\equiv M_z/M$, where $M_z$ is the total planetary mass in metals. 
We assume that the accreted gas envelope $M_{\rm env}=M-M_{\rm core}$ has stellar metallicity $Z_\star$. 
It is possible that the metallicity of the accreted gas is sub-stellar because cores may 
have already sequestered much of the disc's
reservoir of metals. For our nominal 
Jovian-mass
planets, however, only about $\sim$$3\,M_\oplus$ of metals derive from the accreted gas, constituting about 10 per cent of the planet's total $M_z$. Any metal depletion of the accreted gas is therefore a second-order effect. 
Using $M_z=M_{\rm core}+Z_\star M_{\rm env}$, 
we compute the relative metallicity as \citepalias{Thorngren2016}
\begin{equation}\label{eq:z}
\frac{Z}{Z_\star}=1+\frac{M_{\rm core}}{M}\left(\frac{1}{Z_\star}-1\right).
\end{equation}
We assume throughout this work a fixed stellar
metallicity $Z_\star = 10^{-2}$. 
For very massive planets $M\gg M_{\rm core}/Z_\star\gtrsim 3 M_{\rm Jup}$, equation \eqref{eq:z} predicts near stellar metallicity, whereas smaller planets are enriched due to their cores $Z\simeq M_{\rm core}/M$.

\begin{figure*}
	\includegraphics[width=\textwidth]{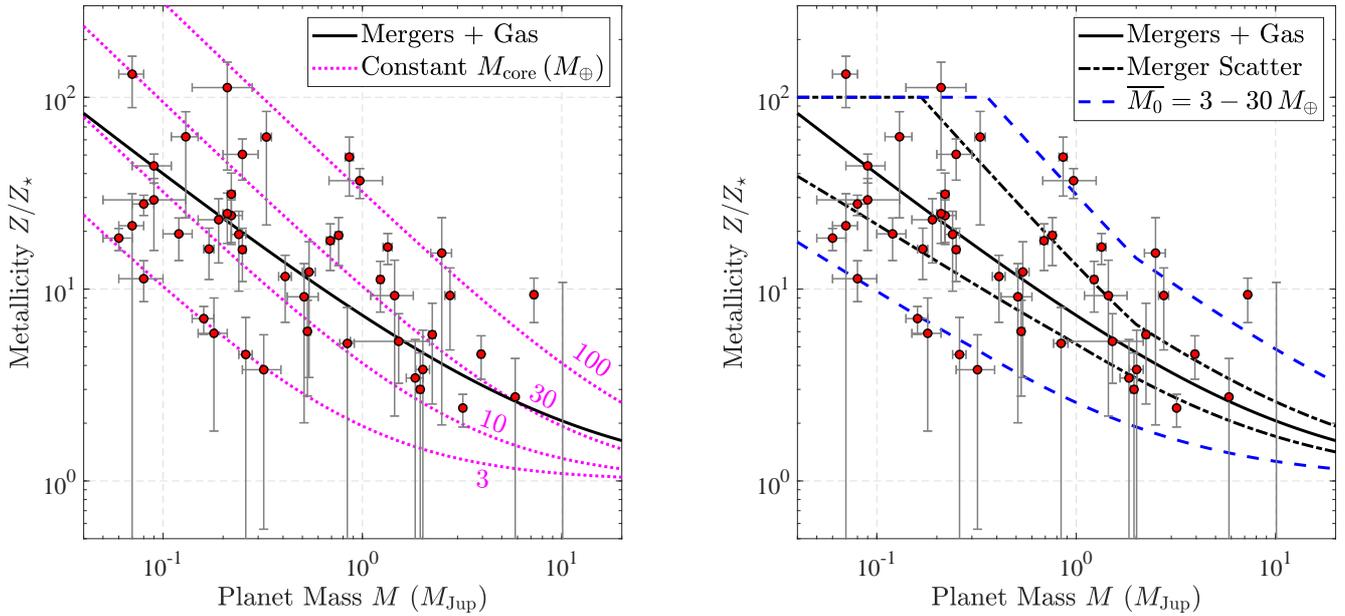}
	\caption{{\it Left:} Planet metallicity, relative to the host star, as a function of planet mass. Observational data points are from \citetalias{Thorngren2016}. Theoretical curves are given by equation \eqref{eq:z} with 
	$Z_\star=10^{-2}$ and using different relations for the core mass $M_{\rm core}(M)$: 
	either constant $M_{\rm core} \propto M^0$ (dotted magenta lines for $M_{\rm core}=3-100\, M_\oplus$),
	or $M_{\rm core}(M)$
	from our theory of concurrent gas accretion 
	and mergers (solid black line; see the corresponding
	solid black line from Fig. \ref{fig:core}) which
	yields a shallower and better fit to the median
	of the data
	--- heavier planets harbour heavier cores. 
	{\it Right:} Same, but showing the theoretical scatter around our nominal model. The dot--dashed black lines 
	bracket the range of metallicity outcomes for an average critical core
	mass of $\overline{M_0} = 10 M_\oplus$, accounting for how mergers are intrinsically discrete and chaotic (see Section \ref{sec:scatter}).
	This intrinsic scatter can account for a significant portion of the observed scatter, even if all systems evolve from practically identical initial conditions. When we vary $\overline{M_0}$ in the range 3--30 $M_\oplus$ (dashed blue lines), all the observations can be accommodated. Metallicities are capped by that of a pure solid core ($Z=1$).}
\label{fig:metallicity}
\end{figure*}

Fig. \ref{fig:metallicity} (left panel) plots $Z/Z_\star$ vs. $M$ and shows that 
concurrent mergers and gas accretion yields a shallower mass--metallicity relation compared to a
conventional 
core-accretion theory in which $M_{\rm core}$ remains
constant during runaway. 
Encouragingly,
the shallower curve appears to fit the mean trend of the observations 
better than the constant $M_{\rm core}=10 M_\oplus$ model.
We turn now to considering what can explain the
large scatter in the observational data.

\subsection{Scatter}\label{sec:scatter}

\citetalias{Thorngren2016} found a large scatter in the metallicity of planets with the same mass, as evident in Fig. \ref{fig:metallicity}. In their deterministic theoretical model, they attributed this scatter to a variation of factor 20 in the disc surface density between different systems.
In our model, the only free parameter is the critical cross-over core mass $M_0$, which depends on the gas opacity and on the initial seed masses and their separations. Presumably, the value of $M_0$ also varies between different discs, producing scatter about our nominal model.

Even for identical discs, however, there is intrinsic
scatter because mergers are inherently stochastic.
There are several components to this intrinsic scatter:
\begin{enumerate}
    \item\label{scatter:t_merge} 
    {\it The distribution of collision times $t_{\rm merge}(M, \Delta a, e)$.} Compact multi-planet systems evolve chaotically, and the time to first collision can change by two orders of magnitude ($\pm 2 \sigma$) for the same planet mass $M$, separation $\Delta a$ between adjacent planets, and initial eccentricities $e$ (fig. 1 in \citealt{Zhou2007}; fig. 1 in \citealt{Rice2018}; figs 3 and 4 in \citealt{HussainTamayo2020}). 
This variation changes the cross-over mass $M_0$ 
where $t_{\rm merge} = t_{\rm cool}$. Since $t_{\rm merge}\propto M^{2\alpha/3}$ (for $M<M_0$) and $t_{\rm cool}\propto M^{-\beta}$, multiplying/dividing $t_{\rm merge}(M)$ by a factor of 10 ($2\sigma$ in each direction) 
decreases/increases $M_0$ 
by a factor of $10^{1/(2\alpha/3+\beta)}\simeq 1.2$.

Most published empirical expressions for $t_{\rm merge}$ are fitted to $N$-body simulations of planets with equal masses and separations. A distribution of unequal masses and separations changes the merger time and potentially increases the scatter \citep{PuWu2015}. 
We defer consideration of these effects to a future study, as it 
requires keeping 
detailed track of the mass and separation distributions. 
\item\label{scatter:e}
{\it The orbital eccentricity $e$.} As before, we use the scaled eccentricity $\tilde{e}\equiv 2ea/\Delta a<1$. 
As long as 
$M<M_0$, the gas disc keeps eccentricities marginally damped between mergers ($t_{\rm damp}\sim t_{\rm merge}$;
Fig. \ref{fig:schematic});
we expect $\tilde{e}$ 
to be limited in range
and to not contribute significantly to
variations in $M_0$ (in contrast to \ref{scatter:t_merge} above).  
The situation changes once $M>M_0$ and $t_{\rm damp}>t_{\rm merge}$, whereupon the disc is unable to damp the planets' eccentricities, which we expect to span a larger
range, up to the orbit-crossing
value of $\tilde{e}=1$.  
Higher initial eccentricities decrease
stability and 
shorten $t_{\rm merge}$. Thus, when calculating
$M_{\rm core}(M)$ using $t_{\rm merge} = t_{\rm cool}$,
we use equation (4) of \citet{Zhou2007} for $t_{\rm merge}$ and allow the 
scaled eccentricity to vary over the full range that they tested, $0\leq\tilde{e}\leq 0.9$.
\item\label{scatter:discrete}
{\it The discrete nature of mergers.} When all planets are of the same mass $M$, each merger can only double the mass of the planet and its core. While the average core mass increases as $M_{\rm core}\propto M^{1/5}$, some planets may be
poised 
just before a doubling, having cores smaller than average for their mass $M$, whereas other planets may 
have just completed 
a doubling, with above-average core masses. In other words, as depicted in Fig. \ref{fig:saw}, the $M_{\rm core}\propto M^{1/5}$ relation is actually the average of a series of `saw teeth' composed of alternating gas accretion and merger episodes. We infer from Fig. \ref{fig:saw} that for a given $M$, the core mass $M_{\rm core}$ can be up to a factor of $2^{2/5}\simeq 1.3$
times 
larger or smaller than average.\footnote{
This is just a rough estimate. 
If all cores reach criticality at the same mass $M_0$, then their 
merger products must be integer multiples of $M_0$;
an ensemble composed of $M_0$ and $2M_0$ cores
would span a factor of $2$ in mass --- this is the two-sided scatter --- larger than
our nominal $2^{4/5}$.
If only one core undergoes runaway gas accretion while the others do not (Section \ref{sec:during}), the two-sided scatter in $M_{\rm core}$ would be $1+M_0/M_{\rm core}$ (each merger adds $M_0$ to the core, without adding any gas). For Jupiter-mass planets,
for which we find $M_{\rm core} \simeq 2M_0$, this
factor is $3/2$,
smaller 
than the nominal $2^{4/5}$.}
Accordingly, we vary $M_{\rm core}(M)$ by this factor in each direction. 

\end{enumerate}

In Fig. \ref{fig:metallicity} (right panel), we show
how much scatter in the mass-metallicity
relation is generated by combining 
the above three effects: \ref{scatter:t_merge} we raise/lower $M_0$ by a
factor of 1.2 from its assumed average value of
$\overline{M_0} = 10 M_\oplus$ 
to account for intrinsic scatter in $t_{\rm merge}$
when $M < M_0$, \ref{scatter:e} we compute a range of
$M_{\rm core}(M)$ relations that account for
a range of pre-merger eccentricities $\tilde{e}$, 
as well as intrinsic scatter in $t_{\rm merge}$
at a given $\tilde{e}$, and \ref{scatter:discrete} we multiply/divide
$M_{\rm core}(M)$ by a factor of $2^{2/5}$
to account for merger discreteness.
The minimum and maximum $M_{\rm core}(M)$
relations so derived
are used to compute the metallicity curves
shown as dot--dashed black lines in Fig. \ref{fig:metallicity}.
For $M\lesssim 2 M_{\rm Jup}$, the scatter is dominated by
the eccentricity effect \ref{scatter:e}, and it is asymmetrical because the nominal curve (solid black line) is for initially circular orbits \citep[at higher masses, merger times are shorter and do not depend as much on the initial $\tilde{e}$; see fig. 1 of][]{Zhou2007}.

As Fig. \ref{fig:metallicity}
demonstrates, even if all protoplanetary discs were identical, the stochasticity inherent in the core's growth through mergers could account for a significant portion of the observational scatter, but not all of it. In order to accommodate the lowest and highest metallicity data, we allow the average critical core mass $\overline{M_0}$ to vary between 3 and 30 $M_\oplus$, and repeat the same calculation as for the nominal $\overline{M_0}=10 M_\oplus$. The results are given by the dashed blue lines in Fig. \ref{fig:metallicity}. By comparing the two panels in Fig. \ref{fig:metallicity}, we see that our initial population of $3-30\,M_\oplus$ cores is transformed by mergers (of varying
effectiveness) into a population of $3-100\,M_\oplus$ cores, a wide enough range to explain the observed scatter
in metallicity.

While our range of cross-over masses $\overline{M_0}$ 
was chosen ad hoc to explain the observations, 
presumably this range reflects real diversity
in protoplanetary discs.
For example, the cross-over mass $M_0$ increases with opacity as $M_0\propto\kappa^{1/(2\alpha/3+\beta)}$ (as can be deduced from Fig. \ref{fig:schematic}, where $t_{\rm merge}\propto M^{2\alpha/3}$ for $M<M_0$, and $t_{\rm cool}\propto \kappa M^{-\beta}$ according to equation \ref{eq:t_cool}). Thus, if some discs are dusty with $\kappa\sim 2\textrm{ cm}^2\textrm{ g}^{-1}$ \citep[as considered by][]{PisoYoudin2014}, their $M_0$ would be higher by a factor of $\simeq$$1.6$.  
A variation in the disc's solid surface density $\Sigma_{\rm solid}$ may further widen the distribution 
of $M_0$, because the isolation masses (before giant impacts begin) scale as $M_{\rm iso}\propto\Sigma_{\rm solid}^{3/2}$ \citep[e.g.][]{KokuboIda2002,Schlichting2014}. 
If we assume that these isolation masses are always separated by a universal $k$ (isolation masses are defined by their Hill radius), then the merger time for $M<M_0$ scales as
$t_{\rm merge}\propto (M/M_{\rm iso})^{2\alpha/3}$. 
By equating this
to $t_{\rm cool}\propto M^{-\beta}$, we find that $M_0\propto M_{\rm iso}^{1/[3\beta/(2\alpha)+1]}\propto\Sigma_{\rm solid}^{5/4}$ for our nominal parameters. We conclude that a factor of $10^{4/5}\simeq 6$ variation in $\Sigma_{\rm solid}$ can underlie the factor of 10 variation we have
invoked in $\overline{M_0}$ to explain the observations,
even if all discs had the same opacity.

\subsection{Heat generated by collisions and atmospheric mass loss}
\begin{figure}
	\includegraphics[width=\columnwidth]{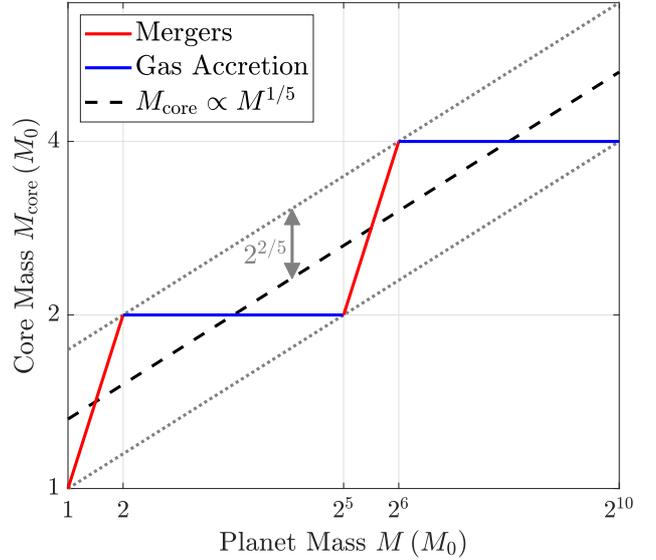}
	\caption{Schematic for how a planet grows by both accreting nebular gas (horizontal solid blue lines) and by merging with its neighbours (sloped solid red lines). In our model, each merger doubles the mass of both the planet and its core, whereas gas accretion does not add mass to the core. 
	Growth by gas accretion destabilizes multi-planet systems ($k\equiv\Delta a/R_{\rm H}$ decreases), eventually leading to a merger which stabilizes the system by increasing $\Delta a$
	and $k$. On average, the core mass grows as $M_{\rm core}\propto M^{1/5}$ (dashed black line, Section \ref{sec:during}), but at any given moment it can be larger or smaller than the average by up to a factor of $2^{2/5}$ (dotted grey lines, Section \ref{sec:scatter}, point \ref{scatter:discrete}).}
\label{fig:saw}
\end{figure}

We have so far assumed that a collision between two planets leads to the merger of both their cores and their envelopes, with no loss of mass. Since the energy released in the collision of two gas giants is comparable to the binding energy of their mutual atmosphere, we may expect that an order-unity fraction of the envelope mass $M_{\rm env}$ is lost in each merger. Nonetheless, as long as the planets are embedded in a  
gas disc, they can re-accrete gas and remain on the $t_{\rm merge}=t_{\rm cool}$ track (Fig. \ref{fig:schematic}). 
If re-accretion is not effective because of gap formation
or disc dispersal (Section \ref{sec:after}),
then envelope loss following a merger will increase
the metallicity, but only by a modest amount.
In the limit $M \simeq M_{\rm env} \gg M_{\rm core}$,
if every merger loses 50 per cent of the envelope mass,
the average $M_{\rm core}(M)$ would equal 
that of a planet of mass $2M$ computed assuming
no mass loss. Replacing $M$ on the right-hand side of equation (\ref{eq:m_core}) with $2M$ shows that
$M_{\rm core}(M)$ would increase by only a factor of
$2^{1/5}$ relative to the case with no gas loss.
Another way to see the effect of mass loss is to examine
Fig. \ref{fig:saw}. Losing 50 per cent of the 
gas with every collision implies that the envelope mass would remain
unchanged while the core mass doubles.
Then the second red `merger' line in Fig. \ref{fig:saw}
would be more nearly vertical, raising $M_{\rm core}(M)$ 
by a factor of $2^{1/5}$ 
above the top dotted grey line.

The release of heat from planetary collisions is
more important during earlier phases of gas giant formation, 
when $M_{\rm env}\ll M_{\rm core}$ \citep{BierstekerSchlichting2019}. When the core dominates the planet's heat capacity, the luminosity of the cooling core can evaporate any existing atmosphere and prevent the accretion of a new one --- planets that formed and were heated by giant impacts might never reach runaway gas accretion. The same concern applies to the formation of sub-Neptunes with voluminous gas atmospheres: if the cores of these planets formed by giant impacts \citep[as suggested by][]{LeeChiang2016}, how did their envelopes manage to cool and accrete?
It might be that this is a problem only for
a small corner of parameter space. 
If a heated core releases its heat 
slowly enough 
\citep[because of inefficient convection; e.g.][]{Staenkovic2012}, the low cooling luminosity would not interfere with gas accretion. If, on the other hand, the core cools fast enough,
it would lose its thermal energy before the gas disc dissipates, and an atmosphere would then be free to accrete \citep[see also the discussion in section 4.2 of][]{Lee2018}. 

\section{Summary and discussion}\label{sec:discussion}

Conventional core accretion theory predicts that gas giants should have heavy-element (`metal') cores of $\sim$$10\,M_\oplus$. 
Cores approaching this mass can, 
within the nebular lifetime, 
undergo runaway gas accretion,
becoming gas-dominated giants \citep{Pollack1996}.
It therefore comes as a surprise to find  exoplanets with $\sim$$100\,M_\oplus$ of metals, constituting more than 30 per cent of their total mass \citep{Thorngren2016,Barragan2018}.

Planetary mergers provide a natural way to create gas giants with especially massive cores.
Each merger fuses two 
cores; 
a series
of mergers can, in principle, produce cores much heavier than $10 M_\oplus$. In this paper we derived a quantitative mass--metallicity relation 
based on this picture, considering mergers
in tandem with planetary gas accretion from the surrounding nebula.
We compared the merger time-scale $t_{\rm merge}$, calibrated from $N$-body simulations \citep{Zhou2007}, with the gas accretion time-scale $t_{\rm cool}$, set by the atmosphere's Kelvin--Helmholtz cooling time \citep{Ikoma2000, Lee2014, PisoYoudin2014, GinzburgChiang2019a}.   
We considered a simplified model of equally spaced, equal-mass planets 
embedded in a gradually depleting gas disc (Fig. \ref{fig:schematic}). As eccentricity damping by the gas disc (i.e. gas dynamical friction) weakens, the initially rocky (i.e. pure metal) planets gravitationally perturb one other onto crossing orbits and merge 
(e.g. \citealt{KominamiIda2002}). 
Eventually, the planets become massive enough to accrete gas from the nebula at the same rate that they merge. After this cross-over mass $M_0\sim 10 M_\oplus$ is reached, planets merge and accrete gas concurrently at equal rates ($t_{\rm merge}=t_{\rm cool}$) that accelerate
with time --- both gas accretion and mergers are in the runaway regime. 
This frenzy of activity occurs 
$\sim$10 au from the central solar-type star,
as it is at these distances that the local gas disc mass can be simultaneously
low enough to permit mergers and high enough
to breed gas giants.

We find that concurrent gas accretion and mergers produces cores whose masses scale as $M_{\rm core}\propto M^{1/5}$, where $M$ is the total mass of the planet (core plus gas envelope). This result is intermediate between growth by gas accretion only ($M_{\rm core} \propto M^0$) and growth by mergers only ($M_{\rm core} \propto M^1$).
Our $M_{\rm core}(M)$ relation yields
a mass--metallicity relation that appears
to fit the median trend observed
by \citetalias{Thorngren2016} well,
as indicated in the left panel of
Fig. \ref{fig:metallicity};
this figure also shows that,
on average, mergers
double or triple the core mass.
More growth is possible because
mergers are inherently stochastic. 
By considering various sources of scatter, we estimated the metallicity range spanned by merged giant planets in the right panel of Fig. \ref{fig:metallicity}. We can reproduce the metallicities of all observed planets in the \citetalias{Thorngren2016} sample 
by positing a range of cross-over core 
masses
$M_0 \sim 3-30\,M_\oplus$.
An appreciable portion of the observational scatter is reproduced even if $M_0$ is
fixed at $10 M_\oplus$ 
--- identical protoplanetary discs with
practically the same initial conditions 
produce a variety of merger
outcomes because compact $N$-body systems 
evolve chaotically.

We emphasize that this paper does not provide an answer as to what sets the 
range of observed final planet masses. We simply assumed that a planet grows by concurrent mergers and gas accretion until it reaches its observed mass; given
the final mass, our
theory computes the range of metallicities arising from this runaway growth phase.
For a discussion of how runaway may end
and how the final masses of gas giants
are determined, see, e.g. \citet{GinzburgChiang2019a} and
\citet{Rosenthal2020}.

\subsection{Observational tests and future work}
What observational signatures might we
expect from a history of mergers?
\begin{enumerate}
\item {\it Higher eccentricities}. 
Mergers require orbit crossing.
For Jupiter-mass planets
spaced $\sim$10 mutual Hill radii
apart, orbit-crossing eccentricities
are on the order of unity. Post-merger
eccentricities will be somewhat lower, as
epicyclic velocities are damped
after an inelastic collision, and there may be 
dynamical friction damping by residual disc gas.
Encouragingly, \citet[][their fig. 13]{Thorngren2016} 
find for their sample of gas giants
that the most eccentric planets 
are the most metal-rich.
\citet{Petigura2017} also report
substantial eccentricities and
metal-enrichment for a sample
of lower-mass sub-Saturns,
and suggest these systems experienced
dynamical instability in the past.

\item {\it Faster spins}. Gravitationally-focused, off-centre
collisions will spin merger products to near break-up
speeds. A correlation between spin rate and core
mass would be a smoking-gun signature of mergers;
for the transiting giant planets studied here,
spin rates may be inferred by measuring the effects
of rotational oblateness on the transit light curve 
\citep{Zhu2014}. 
Caution should be exercised, however, as spin
rates may be slowed by magnetic braking with a residual
gas disc \citep[e.g.][]{Batygin2018,GinzburgChiang2020}.

\item {\it Brighter luminosities}. 
A planet that forms from a major merger
may have a hotter interior
and may therefore radiate for longer
compared to a planet that forms from a series
of minor mergers. In the latter case 
(which asymptotically reduces to formation by smooth
accretion), more of the accretional kinetic energy
is deposited
in a protoplanet's surface layers where it can be
more easily radiated away. A major
merger may trap more of the heat of formation and may
therefore maintain the planet's luminosity for longer
after the disc has dispersed, facilitating
detection by direct imaging.

\item {\it Companions --- or not}. While we have posited that a heavy-metal Jupiter originates from a multi-planet system, it is not clear how many of its companions survive its formation, as all may have been consumed or ejected. If companions do survive, 
planets that are more enriched with metals 
should be separated farther apart. 
Most planets in the \citet{Thorngren2016} sample do not have detected companions, and having a companion does not seem to correlate with the planet's metal mass. 
\citet{Petigura2017} find that sub-Saturns also lack detected companions, which they again interpret as a possible indicator of previous dynamical instability.
\end{enumerate}

We have assumed in our model that planet masses and orbital separations are uniform at all times.
Real merger histories fueled by nebular gas
accretion are inevitably more complex.
We plan to better capture this complexity 
using either Monte Carlo sampling
of published distributions of $t_{\rm merge}$
(e.g. \citealt{PuWu2015}), or direct $N$-body simulations
that prescribe planetary gas accretion (e.g. \citealt{Dawson2015}). Both approaches should provide
better estimates of the scatter in $M_{\rm core}(M)$,
and $N$-body integrations permit tracking of the eccentricity and inclination distributions.

\subsubsection*{Data availability}

No new data were generated or analysed in support of this research.

\section*{Acknowledgements}

We thank Wei Zhu for informative exchanges.
SG is supported by the Heising-Simons Foundation through a 51 Pegasi b Fellowship. 




\bibliographystyle{mnras}
\input{mergers.bbl}




\bsp	
\label{lastpage}
\end{document}